\documentclass[3p,times,procedia]{elsarticle}
\usepackage{nupha_ecrc}
\pdfoutput=1

\volume{00}

\firstpage{1}

\journalname{Nuclear Physics A}

\runauth{}

\jid{nupha}

\jnltitlelogo{Nuclear Physics A}

\usepackage{amssymb}

\usepackage[figuresright]{rotating}

\begin{document}

\begin{frontmatter}



\dochead{XXVIIth International Conference on Ultrarelativistic Nucleus-Nucleus Collisions\\ (Quark Matter 2018)}

\title{Collective effects in nuclear collisions: theory overview}


\author{Jorge Noronha}

\address{Instituto de F\'isica, Universidade de S\~ao Paulo, Rua do Mat\~ao,
1371, Butant\~a, 05508-090, S\~ao Paulo, SP, Brazil}

\begin{abstract}
In these proceedings I review recent developments concerning the hydrodynamic description of the quark-gluon plasma (QGP). I report on the progress towards more realistic simulations and discuss new features about the QGP transport coefficients of the hot (and baryon dense) medium formed in heavy ion collisions. A brief discussion is included on the applicability of hydrodynamics and its extension to the far-from-equilibrium regime.    
\end{abstract}


\end{frontmatter}


\section{Introduction}
\label{section1}

Fluid dynamic behavior can be found at vastly different length scales in nature, ranging from  cosmological scales \cite{largescale}, to kilometer-size scales used in general relativistic hydrodynamic simulations \cite{Baiotti:2016qnr} of neutron star mergers \cite{TheLIGOScientific:2017qsa}, and the length scales corresponding to everyday life phenomena. The ubiquitousness of hydrodynamics stems from its very general core assumptions, i.e., the presence of conservations laws and the assumption of a large separation between different characteristic length scales in the system. This hierarchy is quantified by the Knudsen number $K_N \sim \ell/L$ \cite{CE}, which is the ratio between the relevant microscopic scale $\ell$ and a macroscopic scale $L$ associated with the gradients of conserved quantities. Generally, in the regime where $K_N \ll 1$ one expects fluid dynamics  to emerge. 


The hot and dense quark-gluon plasma formed in heavy ion collisions represents the ultimate frontier of fluid dynamics due to its highly spatially inhomogeneous initial state \cite{Schenke:2012wb}, which leads to ``macroscopic" scales (associated, for instance, with the gradients of energy density) no larger than about a fermi. A simple estimate of the relevant microscopic scale in terms of the inverse temperature $\ell \sim 1/T$ shows that $K_N$ should not be small \cite{Niemi:2014wta,Noronha-Hostler:2015coa}, especially in the collisions of small systems. Of course, collective effects in heavy ion collisions and the answer to questions concerning the thermalization of the quark-gluon plasma, or the lack of it, must stem from many-body effects in quantum chromodynamics (QCD). This illustrates how complicated it is to fully explain the emergence of hydrodynamic behavior in heavy ion collisions, as it will necessarily involve quantifying QCD in the large Knudsen number regime. Despite these conceptual issues, relativistic hydrodynamic models of heavy ion collisions have achieved considerable success in describing the properties of soft hadrons \cite{Heinz:2013th} with impressive predictive power e.g. \cite{Niemi:2015voa,Noronha-Hostler:2015uye}. In fact, hydrodynamic simulations have played a key role in our understanding of the dynamical properties of strongly coupled bulk QCD matter, such as its nearly perfect fluid behavior. In these proceedings I review the recent progress on the hydrodynamic description of nucleus-nucleus collisions based on a personal selection of results that were presented at Quark Matter 2018.          

\section{Recent progress on the hydrodynamical description of the quark-gluon plasma}
\label{section2}

The theoretical description of heavy ion collisions may be roughly divided in three parts: initial state, hydrodynamic evolution, and final state hadronic interactions. The initial state involves the physics that determines the spatial profiles of the energy-momentum tensor $T^{\mu\nu}|_{\tau_0}$ and the currents associated with global symmetries such as the baryon number $J^\mu_B|_{\tau_0}$ at the initial time $\tau_0$. The physics of the initial state of heavy ion collisions is very rich, as was reviewed by A.~Mazeliauskas \cite{aleksas}. Models with gluon saturation such as IP-glasma \cite{Schenke:2012wb} and EKRT \cite{Niemi:2015qia} include the nonlinear gluon dynamics at low transverse momentum that determines the initial multiplicities of each event and the addition of this phenomenon is generally favored by Bayesian analyses of bulk Pb+Pb LHC data \cite{Bernhard:2016tnd}. 

\begin{figure}
\centering
\includegraphics[width=0.9\textwidth]{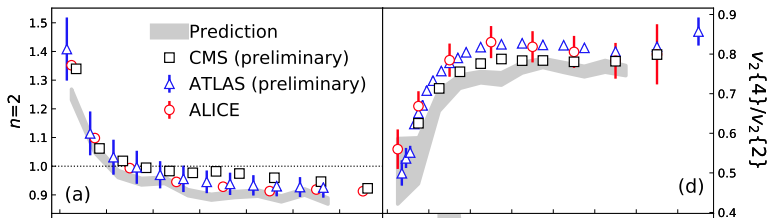}
\caption{Comparison between measurements \cite{Acharya:2018ihu} and hydrodynamic calculations \cite{Giacalone:2017dud} for the ratio between the charged hadron $v_2\{2\}$ in Xe+Xe and Pb+Pb (left) and the ratio of charged hadron cumulants $v_2\{4\}/v_2\{2\}$ (right). This figure was obtained from Ref.\ \cite{matt}.}  
\label{fig1}
\end{figure}

A new development concerning the interplay between nuclear structure and the initial state of heavy ion collisions was revealed in this conference through the recent measurements of anisotropic flow in Xe+Xe collisions performed at the LHC at $\sqrt{s}=5.44$ TeV \cite{Acharya:2018ihu}. As pointed out in \cite{Giacalone:2017dud}, while the ground state of $^{208}$Pb nucleus is spherical $^{129}$Xe is expected to have a moderate prolate deformation and this affects the elliptic flow of the system in central events. In fact, Ref.\ \cite{Giacalone:2017dud} finds that elliptic flow is larger by 25\% in Xe+Xe than in Pb+Pb (at $\sqrt{s}=5.02$ TeV) in the 0-5\% centrality window. This strong enhancement of $v_2\{2\}$ in central Xe+Xe collisions has been seen experimentally, as one can see in Fig.\ \ref{fig1} which also shows the results of theoretical calculations from \cite{Giacalone:2017dud}. Other aspects involving the description of Xe+Xe collisions were explored in \cite{Eskola:2017bup,Niemi:2018ijm}, where it was shown that the ratio $v_2\{2\}\textrm{(Xe+Xe)}/v_2\{2\}\textrm{(Pb+Pb)}$ vs. centrality, while sensitive to the non-spherical deformation of the Xe nucleus, it is not strongly dependent on viscosity. Altogether, these results indicate that hydrodynamic modeling in central heavy ion collision events can be used as a tool to learn about nuclear structure properties. It would be interesting to further explore this idea by considering also the collision of other types of nuclei at LHC.

The hydrodynamic evolution of the quark-gluon plasma in heavy ion collisions is based on Israel and Stewart's formulation of viscous relativistic hydrodynamics \cite{Israel:1979wp}. In this approach, besides the conservation laws of energy and momentum $\nabla_\mu T^{\mu\nu}=0$ and baryon number $\nabla_\mu J_B^\mu=0$, additional relaxation-type equations are used to describe the evolution of the viscous parts of the energy-momentum tensor, i.e., the shear stress tensor $\pi^{\mu\nu}$ and the bulk scalar $\Pi$. A similar relaxation equation for the net baryon diffusion current $q_B^\mu$ must also be added at nonzero baryon chemical potential. To see what these equations look like in their full glory, and how these different dissipative channels couple to each other according to the Boltzmann equation, see Ref.\ \cite{Denicol:2012cn}. At LHC energies most hydrodynamic simulations focus on the mid-rapidity region where the net baryon density is small while at RHIC, especially at the low collision energies in the beam energy scan (BES), this approximation is not justified and effects from the conservation of baryon number (and other QCD conserved currents) must be taken into account. Finally, motivated by recent $\Lambda$ polarization measurements at RHIC \cite{STAR:2017ckg}, great effort has gone into describing the dynamics of vorticity and spin in relativistic hydrodynamics, as reviewed by F.~Becattini \cite{becattini}.

There were some interesting developments concerning both the calculation of transport coefficients and their effects on hydrodynamic simulations. These quantities are of fundamental importance to characterize the near equilibrium dynamics of the quark-gluon plasma and their calculation in QCD from first principles remains a formidable challenge \cite{Ratti:2018ksb}. Progress can be made in some particular limits, for instance at sufficiently large temperatures where the calculation of these coefficients becomes feasible at weak coupling \cite{Arnold:2000dr}. J.~Ghiglieri presented new calculations \cite{Ghiglieri:2018dib,Ghiglieri:2018dgf} valid at next-to-leading order for $\eta/s$ and the shear relaxation time coefficient, $\tau_\pi$. Based on an effective kinetic theory description, a new bound was obtained for the shear relaxation time $\tau_\pi T/(\eta/s) \geq 5$ in the case of ultrarelativistic particles. Also, we note that this bound is violated by nearly a factor of 2 in holographic models of strongly coupled plasmas \cite{Baier:2007ix}, which do not admit a kinetic theory description. This distinction between strong and weak coupling results for $\tau_\pi$ appears in addition to the fact that in weak coupling QCD calculations at very high $T$ $\eta/s \sim 1$ while in holography one obtains the famous KSS result $\eta/s = 1/(4\pi)$ \cite{Kovtun:2004de}. Therefore, this $\tau_\pi$ bound gives important constraints for phenomenological applications because consistency then mandates that in simulations where $\eta/s$ is assumed to be near $1/(4\pi)$ (outside the weak coupling regime) the ratio $\tau_\pi T/(\eta/s)$ should violate the kinetic bound, as in holographic models for the quark-gluon plasma \cite{Finazzo:2014cna}. An overall consistent choice for the values of transport coefficients, contrasting values obtained from weak and strong coupling calculations, should be extremely valuable to elucidate the properties of the QGP formed in heavy ion collisions. In this regard, a step forward was taken in Ref.\ \cite{Dubla:2018czx} where a calculation of $\eta/s$ in QCD from a non-perturbative functional diagrammatic approach was used as input in hydrodynamic simulations. While the temperature dependence of $\eta/s$ obtained in these calculations seemed appropriate, its overall magnitude was not and a downwards shift was needed for a better comparison to data \cite{Dubla:2018czx}. 

\begin{figure}
\centering
\includegraphics[width=0.4\textwidth]{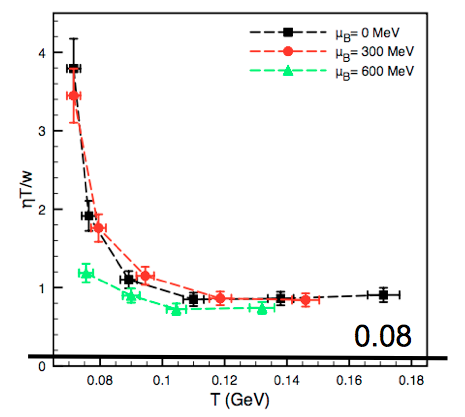}
\caption{Shear viscosity to enthalpy
density ratio vs. temperature for different values of baryon chemical potential in the hadron gas phase computed in Ref.\ \cite{Rose:2017bjz} using SMASH. The solid black line was added in the corresponding figure obtained from J-B.~Rose's talk in this conference \cite{rose}.}
\label{fig2}
\end{figure}

At low temperatures in the hadronic phase, transport coefficients can be estimated using kinetic models describing the interactions among hadrons. Previous calculations \cite{NoronhaHostler:2008ju,NoronhaHostler:2012ug} have shown that the inclusion of heavy resonances is important as one approaches the crossover phase transition and more detailed calculations are certainly needed to determine the temperature (and baryon chemical potential) dependence of transport coefficients in the hadronic phase. J-B.~Rose presented calculations \cite{Rose:2017bjz} for the shear viscosity of a hadron gas obtained using the hadronic code SMASH \cite{hannah}, shown in Fig.\ \ref{fig2}. This plot shows $\eta T/w$, where $w = \varepsilon + P$ is the enthalpy (with $\varepsilon$ being the energy density and $P$ the equilibrium pressure), for a hadron gas as a function of $T$ for different values of baryon chemical potential $\mu_B$. One can see that this quantity is at least an order of magnitude larger than the holographic result (depicted by the solid black line) in the hadron gas phase, becoming very large at low temperatures, while it slightly decreases with increasing $\mu_B$, as expected \cite{Denicol:2013nua}. The fact that $\eta T/w$ assumes a large value and plateaus for a wide range of temperatures in this calculation indicate that this transport coefficient is changing discontinuously in simulations where the hydrodynamic evolution (computed assuming $\eta/s$ to be near the KSS value) is coupled to hadronic transport codes. While such a jump is in principle not forbidden, it is important to keep this in mind when quoting values for $\eta/s$ in heavy ion collisions. 
Within the hydrodynamic model of heavy ion collisions, the only input determined from first principles lattice QCD calculations is the equation of state (EOS), e.g, $P=P(\epsilon)$ at zero $\mu_B$. The vast majority of hydrodynamic simulations performed before this conference employed the equation of state established in Ref.\ \cite{Huovinen:2009yb}, which reflected the state-of-the-art at the time. Since then, different groups have obtained similar results for the QCD EOS with three dynamical quark flavors and calculations for even larger number of dynamical flavors have also been performed \cite{Borsanyi:2016ksw}. The phenomenological consequences of such an update in the equation of state have been discussed in this meeting and it was shown in \cite{Alba:2017hhe} that assumptions regarding the number of active flavors in the EOS (i.e., the inclusion or not of thermalized charm) can have a significant effect in the extraction of $\eta/s$. 

The fact the QGP is not conformal (i.e., $\varepsilon \neq 3P$), especially at the temperatures probed in heavy ion collisions, implies that even if shear effects were negligible (e.g., as in the average expansion of the universe) dissipative corrections would still be present due to a nonzero bulk viscosity $\zeta$. Bulk viscosity was first investigated in event-by-event simulations in \cite{Noronha-Hostler:2013gga,Noronha-Hostler:2014dqa} and the crucial input regarding its importance for the hydrodynamic modeling of heavy ion collisions was made in Ref.\ \cite{Ryu:2015vwa}. There, it was argued that a large value of $\zeta/s$ (with a peak value of $\sim 0.3$) was needed when using IP-Glasma initial conditions, as those produce more radial flow than typical Monte Carlo Glauber models. In this case, bulk viscosity is needed to describe even basic quantities of the evolution such as the mean transverse momentum of produced hadrons. B.~Schenke showed  detailed calculations \cite{Schenke:2018fci} confirming that the relatively compact initial geometry obtained within the IP-Glasma framework is the main ingredient behind the large value of $\zeta/s$ needed to describe the data in these simulations. We note, however, that the highly localized hot spots in the initial energy density profiles obtained from IP-Glasma do not play a significant role in this case, as shown by F.~Grassi \cite{Gardim:2017ruc}. On the other hand, the analysis done by the Duke group \cite{Bernhard:2018hnz} found a significantly small $\zeta/s$ whose magnitude is more comparable to $\eta/s$. One should keep in mind that a meaningful comparison between these calculations is complicated since the latter used Trento initial conditions \cite{Moreland:2014oya} and a different out-of-equilibrium correction $\delta f_{bulk}$ in the freeze-out procedure in comparison to that used in Refs.\ \cite{Ryu:2015vwa,Schenke:2018fci}. The uncertainty in the overall magnitude of $\zeta/s$ has important consequences, as it affects the applicability of models for $\delta f_{bulk}$ at freeze-out \cite{Noronha-Hostler:2013gga} and, if $\zeta/s$ is indeed large near the crossover phase transition, other issues must be taken into account such as the possibility of cavitation \cite{Torrieri:2008ip,Rajagopal:2009yw} (especially in small collision systems). 

\section{Heavy ion collisions at finite baryon density}  

Motivated by the upcoming BES run starting in 2019 at RHIC, part of the focus of the field has shifted towards the description of low energy heavy ion collisions. Besides the inclusion of effects from a conserved baryon current in simulations, other significant changes appear in the presence of nonzero conserved charges. First, strong violations of boost invariance appear as the dynamics of the system at low energies is truly three dimensional \cite{Denicol:2018wdp}. Moreover, in low energy collisions one has to take into account the deceleration of the participating nucleons as the time it takes for the colliding nuclei to pass through each other becomes non negligible, as remarked by C.~Shen \cite{Shen:2018pty}. When it comes to the hydrodynamic evolution, serious uncertainties arise if $\mu_B \neq 0$ as the QCD equation of state cannot be directly computed on the lattice due to the Fermi sign problem \cite{Ratti:2018ksb}. Approximations for the EOS can be made for small $\mu_B/T$ using different techniques on the lattice \cite{Bellwied:2015rza,Bazavov:2017dus}, but it is fair to say that not much is known at large densities. In fact, if QCD displays a critical endpoint (CEP) in the $(T,\mu_B)$ phase diagram in a range accessible to heavy ion collisions \cite{Stephanov:1998dy}, the first step would be to merge what is known from first principles from lattice calculations at $\mu_B=0$ with the expected behavior at the critical point (QCD is in the static universality class of the 3d Ising model) and encode this information in an EOS suitable for implementation in simulations. This has been done by the BEST collaboration in \cite{Parotto:2018pwx}
, as presented by P.~Parotto. 

However, bigger uncertainties appear when considering the out-of-equilibrium behavior of the QGP at large baryon densities. If the QCD CEP is indeed realized, close to this critical regime kinetic theory models (and the hadron resonance gas) should not be used to estimate the value of transport coefficients. As a matter of fact, the baryon diffusion transport coefficient vanishes at the critical point \cite{Son:2004iv}, which can only be obtained in descriptions that correctly incorporate critical behavior. This may be accomplished in holographic models of the QGP that display a critical point, such as those in Refs.\ \cite{Rougemont:2015ona,Rougemont:2017tlu,Critelli:2017oub}. It is not clear at this point how the inclusion of critical behavior in the transport coefficients affects hydrodynamic simulations of the QGP but, in this conference, L.~Du presented recent progress \cite{Du:2018mpf} in this direction by comparing simulations with dynamical sources with a kinetic theory based \cite{Denicol:2018wdp} baryon diffusion with those where the baryon diffusion vanished at the critical point as modeled by holography \cite{Rougemont:2015ona}. Visible differences were found due the inclusion of critical behavior in both the magnitude and in the ÒbumpinessÓ of the solutions. Of course, just the inclusion of critical behavior in the transport coefficients is not enough to describe how hydrodynamic behavior is affected by the presence of critical phenomena. As reviewed in detail by Y.~Yin \cite{yin}, fluctuations become relevant near the critical point and they must be taken into account if the CEP is probed in heavy ion collisions. 

Despite its limitations near the critical point, far from it (though at nonzero baryon density) kinetic theory can still be very useful as it correctly predicts the coupling between the different diffusion currents associated with the QCD conserved charges (i.e, baryon, strangeness, and electric currents), as shown in \cite{Greif:2017byw}.


\section{Far-from-equilibrium hydrodynamics: from paradox to paradigm}
\label{section3}

What is the regime of applicability of relativistic hydrodynamics? When this question was asked during the past century, the answer was straightforward: hydrodynamics is an effective theory defined by a well controlled, systematic expansion around local equilibrium in terms of powers of the Knudsen number \cite{CE}. However, there is now evidence from calculations in both kinetic theory \cite{Denicol:2016bjh,Heller:2016rtz} and holography \cite{Heller:2013fn,Buchel:2016cbj} that such a definition is too simplistic since the gradient series itself diverges. This finding challenges the view that hydrodynamic behavior can be systematically improved to arbitrary order by including more and more powers of gradients. Clearly, this is a relevant issue to the field especially in the description of small collision systems as, experimentally, typical hydrodynamic behavior in the sense of anisotropic flow seems to be present even in these extreme conditions \cite{Aidala:2018mcw}. Furthermore, knowing that the series diverges opens up the possibility of investigating how it may be resummed and what new non-perturbative phenomena in $K_N$ may appear in such a theory (for a review, see \cite{Florkowski:2017olj}).

\begin{figure}
\centering
\includegraphics[width=0.4\textwidth]{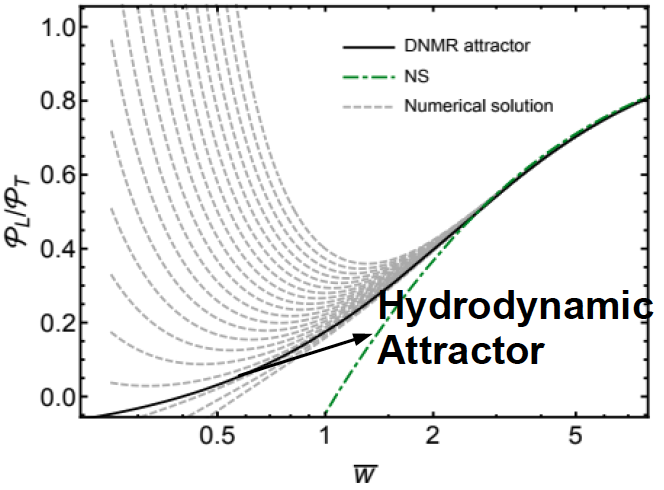}
\caption{Example of hydrodynamic attractor behavior from Ref.\ \cite{Strickland:2017kux}. Different initial conditions (dashed lines) converge to the hydrodynamic attractor (solid line) even though the system is still far from the equilibrium state (which occurs only at very large times $\bar{w}$ when the ratio between longitudinal and transverse pressures goes to unity).}
\label{fig3}
\end{figure}

This new theory, which extends hydrodynamics towards the far-from-equilibrium regime, is currently unknown. However, some of its features have recently emerged with the most prominent being the presence of non-equilibrium attractor behavior, first identified in hydrodynamics in Ref.\ \cite{Heller:2015dha}. Fig.\ \ref{fig3} shows an example of such non-equilibrium structure (solid black) found in \cite{Strickland:2017kux}. This plot shows how the ratio between the longitudinal and transverse pressures in an expanding gas evolves with ``time" $\bar{w}$, for different initial conditions at early times. One can see that the different curves approximately collapse into the attractor even though the system is still far from equilibrium (which is only reached asymptotically when this pressure ratio goes to unity). Several works were dedicated to this rapidly developing topic, see for instance \cite{meiring}. Finally, an interesting possibility to realize far-from-equilibrium hydrodynamic behavior may be to change the starting point of the usual hydrodynamic expansion, choosing an out-of-equilibrium state instead of local equilibrium as done in anisotropic hydrodynamics \cite{Alqahtani:2017mhy}. Progress in this direction was also reported in this conference \cite{McNelis:2018rni}. From the point of view of resummation, anisotropic hydrodynamics performs not only a resummation in Knudsen number but also in the inverse Reynolds number \cite{Denicol:2012cn}, which contributes to the excellent agreement between the results from this approach and exact solutions of the Boltzmann equation \cite{Alqahtani:2017mhy}.   

\section{Conclusions}
\label{section4}

Impressive progress has been achieved over the years in the hydrodynamic description of the QGP. This led to new connections with nuclear structure, established further constraints on the values of QGP transport coefficients, and provided better handling on the challenges that appear in the study of bulk dynamical QCD phenomena. Low energy collisions provide the exciting possibility to measure critical phenomena in a fundamental theory of nature and it is important to remark that even in the absence of the CEP, or if its located beyond the regime covered by heavy ion experiments, low energy heavy ion collisions still provide the unique opportunity to unveil novel out-of-equilibrium phenomena involving the QCD conserved charges. Finally, the study of collective phenomena in heavy ion collisions defines the world's state-of-the-art research in viscous relativistic hydrodynamics, which motivated novel investigations on the foundations of hydrodynamics and its extension to the yet uncharted far-from-equilibrium regime.

\section*{Acknowledgments}

The author thanks the organizers of Quark Matter 2018 for the invitation to give this plenary talk and J.~Noronha-Hostler, W.~A.~Zajc, M.~Luzum, G.~S.~Denicol, J.-F.~Paquet, and H.~Petersen for fruitful discussions. This work was financially supported by
FAPESP under grants 2015/50266-2 and 2017/05685-2
and Conselho Nacional de Desenvolvimento Cient\'ifico e
Tecnol\'ogico (CNPq). 








\end{document}